\documentclass[12pt]{article}
\begin{document}
\begin{center}
\begin{large}
{\bf The island of deformation and shape co-existence in neutron-deficient 
nuclei of the Pb region using relativistic mean field model}
\\[2ex]
\end{large}
M.S. Mehta$^{\dag}$, T.K. Jha$^{\ddag}$, S.K. Patra$^{\S}$ and 
Raj K. Gupta$^{\dag}$
\\
[2ex]
$^{\dag}$ Department of Physics, Panjab University,\\
     Chandigarh 160 014, India\\[2ex]
$^{\ddag}$ P.G. Department of Physics, Sambalpur University,\\
     Jyoti Vihar, Burla 768 019, India\\[2ex]
$^{\S}$ Institute of Physics, Sachivalaya Marg, Bhubaneswar 751 005, India.
\\
[2ex]
\end{center}

\bigskip
\begin{small}
\centerline {\bf Abstract:}
We have investigated the ground-state structures of even-even 
neutron-deficient isotopes of Hg and Pb nuclei within the framework of a 
deformed relativistic mean field formalism for a number of commonly used 
parameter sets, namely NL1, NL3, NL$-$SH, NL$-$RA1 and TM1. The ground 
states of a bunch of Hg and Pb-isotopes towards the proton-dripline are found 
to be deformed for all the forces, with a constant pairing gap. The small 
differences in the ground- and the first-excited state binding energies 
predict a sea of low-lying excited states, and hence of shape co-existence, 
for both the Hg and Pb nuclei. In general, a discrepancy between the 
experimentally observed and theoretically predicted (sign of) quadrupole 
deformations is noticed in this mass region. The constrained potential 
energy surfaces and the single-particle energy spectra analyzed for some 
selected nuclei show that the known large shell gap for Pb nuclei at Z=82 
is almost extinguished for its proton-rich isotopes.

\bigskip
\bigskip
\bigskip
\end{small}

\noindent{\bf PACS: 21.10.Dr, 21.10.Tg, 21.60.-n, 21.60.Fw}\\

\noindent{{\bf Keywords}: Drip-line nuclei, binding energies, quadrupole deformations, 
single-particle energy spectra, relativistic mean field theory}\\

\vfill
\eject
\baselineskip 18pt

\section {Introduction}

The study of the stuctrures of Hg and Pb nuclei, in particular of the 
neutron-deficient isotopes, has been a puzzle and still remains a puzzle. 
A variety of structures have been proposed by different calculations, some 
questioning even the magicity of Z=82 shell for nuclei at the 
proton-drip line. Already in early 1970's, based on the Hartree-Fock (HF) 
calculation using Skyrme force, a shape transition from oblate to prolate 
in the ground-state of neutron-deficient Hg isotopes was reported with the 
decrease of mass number \cite{faessler72,calliau73,kolb75}. On the contrary, 
the later calculations based on the Strutinsky prescription with BCS pairing 
\cite{bengtsson87,nazare93} predicted an oblate ground-state for 
neutron-deficient Hg nuclei, agreeing with the experimental data on first 
excited 2$^+$ states. 

The relativistic mean field (RMF) calculation for the neutron-deficient Hg 
and Pb nuclei was first made by one of us and collaborators in 1994 
\cite{patra94,yoshida94}, using the NL1 parameter set with the pairing 
interaction in BCS formalism, of constant gap parameters as in 
\cite{moeller88}. This calculation gave rather unexpected results. 
The ground-state shape of even-even $^{178,182,184,186}$Hg was predicted 
to be prolate, and that of $^{184-196}$Pb isotopes as deformed having 
co-existing prolate and oblate shapes. A further stricking result was the 
superdeformed (SD) ground-state in $^{180}$Hg. According to a critique by 
Heyde et al. \cite{heyde96}, the above RMF results must be due to a 
particular choice of the force parameter and the very schematic treatment 
of pairing. In response to this comment, Takigawa et al. \cite{taki96} 
showed that a reduction of the gap parameter for neutrons by a factor of 
two with NL1 force, or by changing the force to NL-SH set, the ground-state 
of $^{180}$Hg does indeed changes from the superdeformed to one of the 
normal deformation. Depending on the strength of pairing gap for neutrons, 
it shifts from oblate to prolate, the SD state always remaing an excited 
state. However, for other neutron-deficient Hg and Pb nuclei, the change 
of the parameter set from NL1 to NL-SH did not alter much the predictions 
of either the prolate deformed ground-states of Hg or the deformed structures 
for Pb nuclei. However, another RMF calculation using NL3 force with BCS 
pairing shows all the neutron-deficient Pb nuclei to be spherical 
\cite{lala99}. This means that the criticism of Heyde et al. \cite{heyde96} 
for the RMF predictions of Hg and Pb ground-state strutures is answered only 
partially and the question of checking the sensitivity of RMF calculations 
to the choice of force parameter and/or the strength of pairing force needed 
further investigation. 

Some further effort has been made for understanding the role of pairing 
strength in RMF calculations. Yoshida and Takigawa \cite{yoshida97} have 
shown that a constant strength of the pairing interaction G (which makes 
the gap parameters strongly deformation dependent), instead of constant 
gap parameters (independent of the deformations) with NL1 force itself 
reproduces the oblate shapes and charge radii of neutron-deficient Hg 
isotopes and restores the magicity of Z=82 shell for all the Pb isotopes. 
In addition, the isotope shifts of Pb isotopes are well reproduced. This 
makes one feel that the problem is settled once for all, if the results 
of this calculation will be independent of the choice of the force parameter 
set NL1 and the average gap patrameters used for solving the pairing 
interaction equations for G. Instead, these authors suggested that, for 
exotic nuclei, the alternative, deformed relativistic Hartree-Bugoliubov 
(RHB) approach is superior to the BCS theory used by them, since it gives 
a unified description of both the mean-field and pairing correlations. 
Such a study is also taken up very recently by Nik\v si\'c et al. 
\cite{niksic02} for the ground state properties of neutron-deficient Pt, 
Hg and Pb nuclei, using the NL3 parameter set with finite range Gogny 
interaction DIS \cite{berger84}. However, the results of this calculation 
opens up the problem once again, since they again predict $^{188-194}$Pb 
nuclei as oblate deformed in their ground-states. Thus, in order to keep 
the magicity of Z=82 shell, a new parametrization, NL-SC, is proposed which 
takes into account the sizes of spherical shell gaps, particularly at Z=82, 
and hence reproduces the experimental data on binging energies, radii and 
ground-state deformations of these nuclei. However, such an effective 
interaction could not be considered as general and may not describe 
accurately the ground-state shapes of nuclei in other regions of shape 
co-existence.

The puzzle on structures of Hg and Pb nuclei at the proton-drip line is further complicated by 
the most recent large-scale mass measurements of proton-rich nuclei around Z=82 
\cite{novikov02}. The reduction of two-proton separation energies [rather, the differences 
between two-proton separation energies of adjacent even-even nuclei with the same neutron 
number, $\delta _{2p}=S_{2p}(Z,N)-S_{2p}(Z+2,N)$], for Pb nuclei with N$\approx$106 
\cite{novikov02}, combined with the systematics of older measurements of $Q_{\alpha}$-values 
and $\alpha$-reduced widths \cite{toth99}, suggests the weakening of the spherical Z=82 shell 
for neutron number in between N=82 and 126. Using the non-relativistic self-consistent 
mean-field theory with Skyrme interaction or the RMF model with NL3 and NL-Z2 parameter sets 
in BCS approximation of pairing, Bender et al. \cite{bender02} show that the systematics of 
$\delta_{2p}$ could well be described quantitatively in terms of the deformed ground states of 
Hg and Po isotopes. Already, a tripple shape co-existence (spherical, oblate and prolate shapes 
of almost identical excitation energies) is observed in $^{186}$Pb \cite{andreyev00}. Also,
the earlier calculations based on RMF approach show that the Z=N=28 looses its magicity for 
nuclei approaching the neutron dripline \cite{patra97,warner96} and that for the valley of 
superheavy nuclei the sequence of magic numbers is different from the usual one \cite{patra00}. 
These results make us believe that we could still learn much from the standard RMF model 
calculations, using the BCS formalism for pairing interaction, provided a systematic attempt 
is made for a number of parameter sets. We do this here for the parameter sets NL1, NL3, NL-SH, 
NL-RA1 and TM1, for the pairing gap parameters still taken from Madland and Nix 
\cite{moeller88}. This covers almost the entire range of the generally used force parameter 
sets for RMF model calculations. The case of zero pairing is also investigated. We find 
that some of the results do depend strongly on the choice of the parameter set. The already
used parameter sets NL1, NL3 and NL-SH are also included here for the relative comparisons of
the calculations made under similar conditions.

The paper is organised as follows: In section 2 we outline the essential formalism for the 
relativistic mean field Lagrangian. The results of our calculation for $^{170-200}$Hg 
and $^{178-208}$Pb nuclei are discussed in section 3. We look for the shape co-existence and 
possible deformation effects in this island of Z=80 and 82 nuclei. The ground-state and the 
first exited-state solutions are obtained for all the nuclei studied and the constrined 
potential energy sufaces (PES) are calculated for a few illustrative nuclei. For the deformation 
effects, the quadrupole deformation parameters and the single-particle energy spectra for both 
the spherical and deformed nuclei are analyzed. Finally a summary of our conclusions is added 
in section 4.

\section {The formalism}

In the last two decades, the RMF model has become known to be a very powerful tool to explain 
the properties of finite nuclei and infinite nuclear matter \cite{machleidt89,patra91,patra01}. 
The RMF method has the advantage that, with proper relativistic kinematics and with the mesons 
and their properties already known or fixed from the proprties of a few nuclei 
\cite{patra91,rufa88,gambhir90}, the method gives excellent results for binding energies, 
root mean square (rms) radii, quadrupole and hexadecupole deformations and other nuclear 
properties, not only of spherical, but also of deformed nuclei. One of the major attractive 
features of the RMF formalism is that the spin-orbit strength and associated nuclear shell 
structure automatically arise from meson-nucleon interaction \cite{horowitz81,serot86}.

We start with the relativistic Lagrangian density for a nucleon-meson many-body system,
\begin{eqnarray}
{\cal L}&=&\overline{\psi_{i}}\{i\gamma^{\mu}
\partial_{\mu}-M\}\psi_{i}
+{1\over{2}}\partial^{\mu}\sigma\partial_{\mu}\sigma
-{1\over{2}}m_{\sigma}^{2}\sigma^{2}-{1\over{3}}g_{2}\sigma^{3}
-{1\over{4}}g_{3}\sigma^{4}
-g_{s}\overline{\psi_{i}}\psi_{i}\sigma\nonumber\\
&-&{1\over{4}}\Omega^{\mu\nu}\Omega_{\mu\nu}
+{1\over{2}}m_{w}^{2}V^{\mu}V_{\mu}
+{1\over{4}}c_3(V_{\mu}V^{\mu})^2-g_{w}\overline\psi_{i} \gamma^{\mu}\psi_{i}
V_{\mu}-{1\over{4}}\vec{B}^{\mu\nu}.\vec{B}_{\mu\nu}\nonumber\\
&+&{1\over{2}}m_{\rho}^{2}{\vec R^{\mu}}.{\vec{R}_{\mu}}
-g_{\rho}\overline\psi_{i}\gamma^{\mu}\vec{\tau}\psi_{i}.\vec
{R^{\mu}}-{1\over{4}}F^{\mu\nu}F_{\mu\nu}-e\overline\psi_{i}
\gamma^{\mu}{\left(1-\tau_{3i}\right)\over{2}}\psi_{i}A_{\mu}.
\end{eqnarray}
The field for the $\sigma$-meson is denoted by $\sigma$, that for the $\omega$-meson by 
$V_{\mu}$ and for the isovector $\rho$-meson by $\vec R_{\mu}$. $A^{\mu}$ denotes the 
electromagnetic field. The $\psi_i$ are the Dirac spinors for the nucleons whose third 
component of isospin is denoted by $\tau_{3i}$. Here $g_{s}$, $g_{w}$, $g_{\rho}$ and
${e^{2}\over{4\pi}}={1\over{137}}$ are the coupling constants for $\sigma$, $\omega$, 
$\rho$ mesons and photon, respectively. $g_2$, $g_3$ and $c_3$ are the parameters for the 
nonlinear terms of $\sigma$- and $\omega$-mesons. M is the mass of the nucleon and 
$m_{\sigma}$, $m_{\omega}$ and $m_{\rho}$ are the masses of the $\sigma$, $\omega$ and 
$\rho$-mesons, respectively. $\Omega^{\mu\nu}$, $\vec{B}^{\mu\nu}$ and $F^{\mu\nu}$ are the 
field tensors for the $V^{\mu}$, $\vec{R}^{\mu}$ and the photon fields, respectively
\cite{gambhir90}.

From the relativistic Lagrangian we get the field equations for the nucleons and mesons. 
These equations are solved by expanding the upper and lower components of Dirac spinors and 
the Boson fields in a deformed harmonic oscillator basis with an initial deformation. 
The set of coupled equations is solved numerically by a self-consistent iteration method. 
The centre-of-mass motion is estimated by the usual harmonic oscillator formula
$E_{c.m.}=\frac{3}{4}(41A^{-1/3})$. The quadrupole deformation parameter $\beta_2$ is 
evaluated from the resulting quadrupole moment \cite{gambhir90} using the formula,
\begin{eqnarray}
Q=Q_n+Q_p={\sqrt{{9\over{5\pi}}}}AR^2{\beta_{2}},
\end{eqnarray}
where $R=1.2A^{1/3}$. The total binding energy of the system is,
\begin{eqnarray}
E_{total}=E_{part}+E_{\sigma}+E_{\omega}+E_{\rho}+E_c+E_{pair}+E_{c.m.},
\end{eqnarray}
where $E_{part}$ is the sum of the single-particle energies of the nucleons and $E_{\sigma}$, 
$E_{\omega}$, $E_{\rho}$, $E_c$ and $E_{pair}$ are the contributions of the mesons fields,
the Coulomb field and the pairing energy, respectively. For the open shell nuclei, the effect 
of pairing interactions is added in the BCS formalism. The pairing gaps for proton 
$\triangle_p$ and neutron $\triangle_n$ are calculated from the relations \cite{moeller88},
\begin{eqnarray}
{\Delta}_{p}&=&r b_sZ^{-1/3}e^{(sI-tI^2)}\nonumber\\
{\Delta}_{n}&=&r b_sN^{-1/3}e^{-(sI+tI^2)}
\end{eqnarray}
where $r=5.72 MeV$, $s=0.118$, $t=8.12$, $b_s=1$ and $I=(N-Z)/(N+Z)$.

\section {Results and Discussions}
In this section, we discuss the results of our relativistic mean field 
calculations. The commonly used NL1 \cite{rein86}, NL$-$SH \cite{sharma93}, 
NL3 \cite{lala97,sharma00}, TM1 \cite{toki} and the more recent NL$-$RA1 
\cite{rash01} parameter sets are employed for the present calculations. For 
the NL3 parameter set, we have used the numerical data from \cite{sharma00}, 
which is slightly different (in decimal places) from that given in their 
earlier work \cite{lala97} and used in later work \cite{lala99} mentioned 
in the Introduction; this rounding off of the numbers makes a difference of 
about 1 MeV in the total binding energy. We display our results in Table 1 
for only some Hg and Pb isotopes, where the experimental data 
(on both the binding energy and deformation parameter) are available, and 
discuss the complete calculations for $^{170-200}$Hg and $^{178-208}$Pb nuclei 
in Figs. 1-5.

In Table 1, we notice that the calculated binding energies BE and quadrupole 
deformation parameters $\beta_2$, for both the cases of with and without 
pairing, are quite close to the experimental values, taken from Refs. 
\cite{raman87,audi97}. Also, interesting enough, all the parameter sets give 
almost similar results. A closer inspection of Table 1 shows that NL3 and 
NL-RA1 sets are somewhat better than the other parameter sets for predicting 
the experimental binding energies,  NL3 being superior in some cases and 
NL-RA1 in some other cases. Note that the evaluation of quadrupole deformation 
parameter ${\beta_2}$ from the experimental B(E2)-value \cite{audi97} gives 
only the absolute ${\beta_2}$-value and hence the comparison between the calculated 
${\beta_2}$ and experimental ${\beta_2}$ is limited in this respect of the sign.

In Fig. 1, the difference $\Delta E$ of the binding energies between the
intrinsic ground-state and the intrinsic first excited-state (from here onwards, 
refered to simply as the ground-state and excited-state) is shown for
Hg and Pb isotopes. The upper and lower panels show without and with pairing 
interaction, respectively. Almost all the Hg isotopes (without pairing, upper panel) 
show very small energy difference between the ground- and first excited-states,
i.e. without pairing effects, $\Delta E$ is within $\sim$1.5 MeV for almost all the 
Hg nuclei with all the parameter sets used. The main exception is $^{184}$Hg, where
$\Delta E{\approx}$3.5 MeV for almost all the parameter sets. The other
notable cases are $^{186}$Hg for NL-SH, $^{194,196}$Hg for NL1 and $^{188}$Hg
for TM1 force, where $\Delta E>$2 MeV. Hence, in general, for the Hg nuclei
considered here, we could say that the first excited-state is not far from
the ground-state, a result very much consistent with the earlier theoreticel
predictions \cite{faessler72,frauendorf75,dracolis88}. Similarly, in the case
of Pb isotopes (without pairing, upper panel), the difference $\Delta E$ is also 
within ${\approx}$1.5 MeV for most of the nuclei, in particular for $^{178-192}$Pb
and $^{200,202,206,208}$Pb nuclei for all the parameter sets. The noteable exception 
is the TM1 fore for $^{178,180}$Pb and $^{200-208}$Pb where $\Delta E$ is much
larger. For $^{178-188}$Pb and $^{206,208}$Pb, the $\Delta E$ is very small, 
$\sim$0.8 MeV or less. In $^{194,196}$Pb nuclei, the difference $\Delta E$ is more 
than 1.5 MeV for almost all the parameter sets. Thus, once again, with a few 
exceptions, the ground-states and the first excited-states for the Pb nuclei
investigated here are nearly degenerate. In other words, with zero (or week) pairing 
there are low-lying excited-state solutions or shape co-existing states in the 
island of Hg and Pb nuclei.

In the lower panel we plot the same $\Delta E$, but for including the pairing
effects. A comparison between two (upper and lower) panels allows us to see the 
influence of pairing correlations (note the difference in the ordinate scales 
of the two panels). In general terms, we notice that the pairing
interaction has very little effect on the ground- and excited-state energies
in Hg and Pb nuclei. For example, $^{170-180,190,198,200}$Hg and nearly all Pb 
nuclei, for most of the forces, have the energy difference $\Delta E$ less than 1.5 MeV.
The notable exceptions are $^{182-188}$Hg and $^{194-198}$Pb isotopes, which
also have $\Delta E {\leq}$2.7 MeV, for many parameter sets.

In order to get a better realization of the barrier between the ground-
and excited-states, we have plotted the potential energy surfaces (PES) in Fig. 2
for two illustrative, strongly deformed nuclei, $^{180}$Hg and $^{186}$Pb. 
This is a constrained calculation where, instead of minimizing $<H_{0}>$, we have 
minimised $<H-{\lambda}Q>$, with ${\lambda}$ as a Lagrange multiplier and Q, the 
quadrupole moment \cite{patra02}. This gives us the binding energy as a function of the
quadrupole deformation parameter ${\beta_{2}}$. In Fig. 2, we observe that
there is a low-lying solution near to the ground-state for both $^{180}$Hg
and $^{186}$Pb nuclei. The two solutions have about the same energy difference of 
${\sim}$3 MeV for both $^{180}$Hg and $^{186}$Pb nuclei, and for all the parameter 
sets. Thus, the shape co-existence seems to be nearly independent of the chosen 
parameter sets. It may be noted that the (multiple) shape co-existence is a very 
well studied phenomenon in Pb nuclei \cite{julin01} and, as already mentioned above, 
with $^{186}$Pb as the best studied case, having three differently shaped $0^+$ states 
observed recently \cite{andreyev00}. However, the present RMF calculations do not give 
the observed spherical solution for $^{186}$Pb nucleus.

Fig. 3 gives the variation of the quadrupole deformation parameter ${\beta_{2}}$ as a 
function of mass number A, for the two cases of, respectively, without and with 
(upper and lower panel) pairing interactions. In the upper panel (without pairing) for 
Hg-isotopes all forces show the same transition of shape with the mass number. This
transition is nearly independent of adding the pairing strength (compare the upper and 
lower panels for Hg nuclei). Thus,  for Hg nuclei, we notice no significant change 
for adding the pairing interaction, but for Pb nuclei, for all the parameter sets, 
we observe that the pairing has a very significant effect on their ground state 
structures. The oblate shape for $^{182-190}$Pb nuclei becomes prolate when pairing is 
added. This happens because the first excited-states are very close to the ground-states 
for these nuclei (see Fig. 1) and a small change in the input parameters, like for the 
pairing strength, can flip the ground-state solution into an excited-state or vice 
versa. 

For Hg nuclei, almost independendent of pairing strength, the maximum (prolate) 
quadrupole deformation is for $^{184}$Hg (${\beta_{2}}$=0.30-0.33 for the five 
parameter sets used), and for $^{170-176}$Hg, the the ground-state shape is spherical 
(near spherical) with all the forces. The change in shape from prolate to oblate 
takes place at mass number A=188 for NL1, at A=190 for the other three parameter 
sets, while TM1 shows this change a still further at A=192. In other words, 
the $^{178-188}$Hg nuclei are prolate deformed for all the parameter sets used. 
In the experiments \cite{dracolis88}, however, the estimated prolate-oblate energy 
differences for $^{184-188}$Hg nuclei allow these authors to conclude that the lower 
of the co-existence states is an oblate one and the other low-lying stable prolate 
solution is the excited state, supporting the earlier theoretical results of Bengtsson 
et al. \cite{bengtsson87} using Woods-Saxon potential and another RMF calculation 
\cite{lala99} using the (older) NL3 parameter set of Ref. \cite{lala97}. Note, however, 
that we have used here the more recent NL3 parameter constants of Ref. \cite{sharma00} 
and, as already mentioned above, the results of two NL3 parameter sets (old and recent) 
could differ by about 1 MeV, which is of the order of $\Delta E$ for some of these nuclei 
(see Fig. 1) and hence may change the prolate solution to oblate one. On the other hand, 
for Pb nuclei, with pairing effects included (lower panel), we notice a change from 
spherical to oblate deformation, right at the begining for $^{180}$Pb or $^{182}$Pb, 
depending on the parameter set used, and again from oblate to prolate region at $^{180}$Pb 
or $^{182}$Pb. The nuclei $^{182-190}$Pb are prolate for all sets. Alternatively, with 
pairing effects switched off (upper panel), almost all the Pb nuclei studied here are
oblate deformed with $\beta\sim$0.2. In particular, our predictions for $^{192,194}$Pb, 
e.g. ${\beta_2}$=-0.172 and -0.166 for NL3 set, are in excellent agreement with the 
experimental data (${\beta_2}$=-0.18, -0.175 and -0.17 \cite{dendoo89}) where the 
sign of ${\beta_2}$ is determined on the basis of a theoretical calculation 
\cite{bengt87a}. However, we consider the neutron deficient $^{182-190}$Pb as prolate
deformed, advocating strong pairing correlations, since the neghbouring $^{178-188}$Hg 
nuclei are prolate deformed, independent of pairing as well as parameters of the forces 
used.

Next, Figs. 4 and 5 (with and without pairing, respectively) display the single-particle 
spectra for two illustrative nuclei $^{186}$Pb and $^{208}$Pb, i.e. the most deformed 
and the most spherical isotopes of the chosen series, for the representative NL1 and NL3 
parameter sets. The results obtained for the other parameter sets NL$-$SH, NL$-$RA1 and 
TM1 are similar. From Fig. 4 (where pairing effects are included), we notice that the 
shell gap for the spherical nucleus $^{208}$Pb is well distinguished at Z=82 (and at N=126), 
which in the deformed nucleus $^{186}$Pb seems to have been reduced considerably, both 
for protons and neutrons (more so for protons). All the four parameter sets show 
almost the same result. We have also repeated our analysis for the case of not 
including the pairing interactions (Fig. 5) and found that there is no appreciable change 
in the shell gaps, except that for the deformed $^{186}$Pb nucleus the gap at Z=82 is now 
larger ${\sim}$2 MeV (for NL1 force), to be compared with ${\sim}$4 MeV for the spherical 
$^{208}$Pb nucleus. Thus, we can say that some light, neutron-deficient isotopes of 
Pb are deformed and show the signatures of gap reduction, as is observed in the 
experiments of Toth et al. \cite{toth99}. It may be noted here that if the single 
particle energy shell gap near the magic shell is larger than the two-nucleon 
interaction strength, then the shell gap at the closed shell has to be considered. 
On the other hand, for a shell gap smaller than the two-nucleon interaction matrix, the 
considered nucleon can jump from this shell to the higher orbits.

Finally, a few words about the co-existing SD state as the ground-state, predicted in Refs.
\cite{patra94,yoshida94} for $^{180}$Hg using NL1 parameter set with full BCS pairing gap 
parameter from Ref. \cite{moeller88}. In order to verify this result, we have repeated this 
calculation by using NL1, NL-SH, NL3 and NL-RA1 parameter sets by increasing the oscilator 
shell number to $N_F=N_B$=16. The shapes of both the normal and super-deformed solutions 
are investigated, by taking the cases of both with and without pairing interactions into 
account. For the NL1 parameter set with pairing, the binding energy BE=1420.420 MeV 
(at ${\beta_2}$=0.569) for the SD solution, and the BE=1422.405 MeV for normal deformation 
(${\beta_2}=0.323$). For the case of without pairing, the binding energy (and deformation 
parameter) for the normal and SD configurations are BE=1420.304 ($\beta_2=0.328$) and 
BE=1420.420 MeV ($\beta_2=0.569$), respectively. This means that in $^{180}$Hg for NL1 
parameter set, the SD solution is very close to the ground-state and could be the 
ground-state solution itself, as was suggested by the earlier calculations 
\cite{patra94,yoshida94} based on NL1 parameter set. However, for the other three 
parameter sets we notice in Table 2 that the normal and SD solutions are rather far apart. Hence, the NL1 parameter set seems to have 
the special feature of predicting a shape co-existing super-deformed state with the 
ground-state, contrary to other parameter sets NL-SH, NL3 and NL-RA1 studied here. 

\section {Summary and Conclusions}
In the present investigation we have calculated the binding energy, the single-particle 
energy spectrum and the quadrupole deformation parameter ${\beta_{2}}$ for some 
neutron-deficient Hg and Pb isotopes, using the deformed relativistic mean field 
formalism with the commonly employed parameter sets NL1, NL3, NL$-$SH, TM1 and NL$-$RA1. 
We find a low-lying excited-state in most of the chosen nuclei, for all the five 
parameter sets. Thus, the shape co-existance in neutron-deficient Hg and Pb nuclei is 
predicted to be nearly independent of the force parameters and pairing correlations. Also, 
clear deformed structures for the neutron-deficient $^{178-188}$Hg and $^{182-190}$Pb 
isotopes are noticed, irrespective of the chosen parameter set, due to melting away of the
usual Z=82 shell closure. 

For the binding energy comparisons with experiments, the NL3 and NL$-$RA1 parameters have 
a slight edge over the other parameter sets. The potential energy surfaces are investigated 
with all the five forces by taking the extremely deformed $^{180}$Hg and $^{186}$Pb nuclei 
as the representative cases. Both the nuclei show a similar behavior for each parameter set. 
The single-particle energy spectrum shows a large shell gap at Z=82 (and also at N=126) for 
the most spherical nucleus $^{208}$Pb, but the same is not true for the most deformed 
$^{186}$Pb. For the deformed nucleus the shell gap at Z=82 is reduced considerably. 

Concluding, we emphasize that a bunch of Hg and Pb nuclei in the considered island of Hg and Pb
nuclei are deformed and that the magic character at Z=82 is lost for the neutron-deficient 
deformed Pb nuclei. The discrepancy between the deformed RMF predictions and the experiments 
for the quadrupole deformation is due to the low-lying excited states (shape coexistence)
obtained in these nuclei. Apparently, to reproduce the experimental data, it is essential to
perform a more careful calculation.\\ \\

\par\noindent
{\bf Acknowledgements}\\
We thank Prof. A. Ansari for a careful reading of the manuscript and for many valuable 
suggestions. One of us (RKG) is grateful to the Department of Science and Technology, 
Govt. of India for support of this research work in terms of a Sr. Research Scientistship
to him. MSM and TKJ thank the Institute of Physics, Bhubaneswar, for the kind hospitality 
during the completion of this work. 

\vfil
\eject
\centerline {\bf Figure Captions}
\bigskip
\bigskip
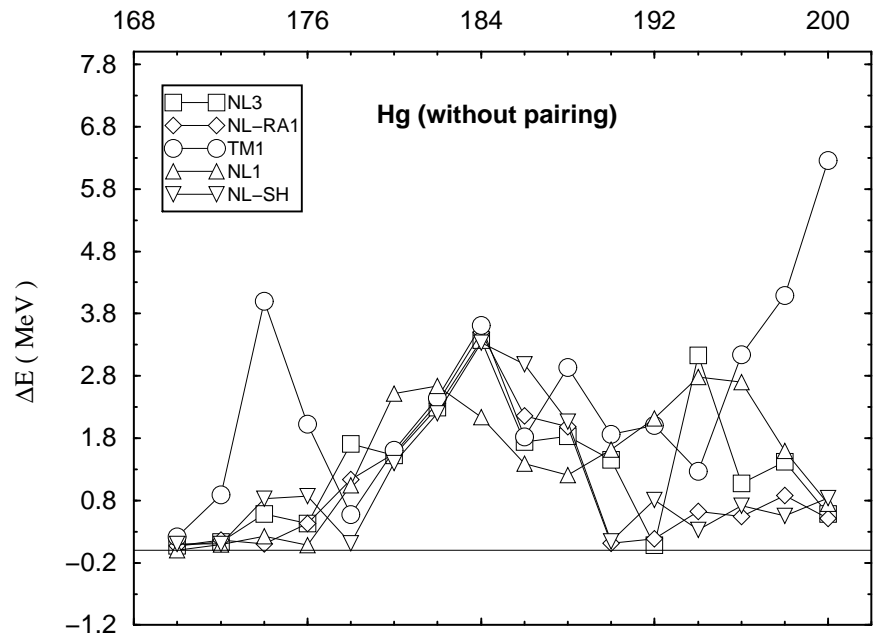
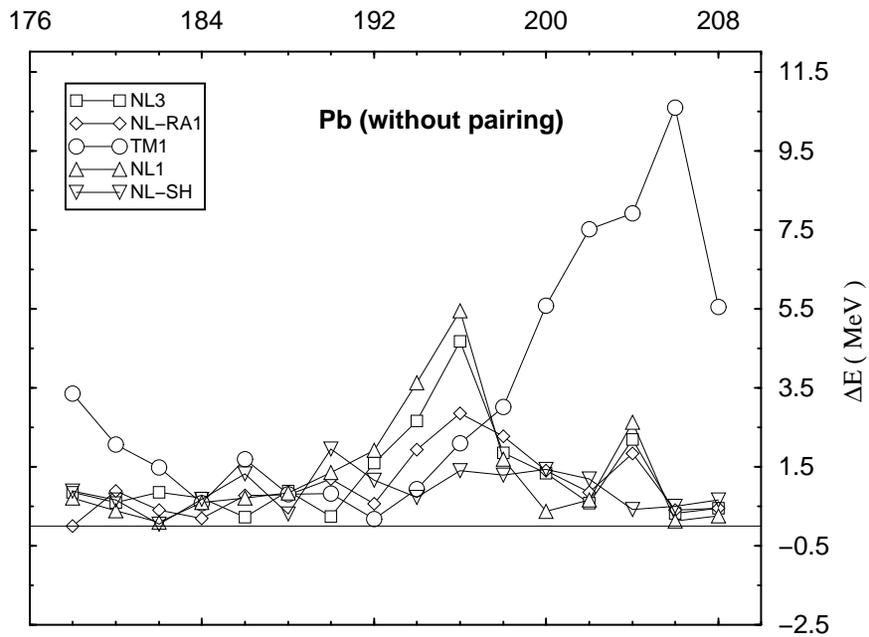
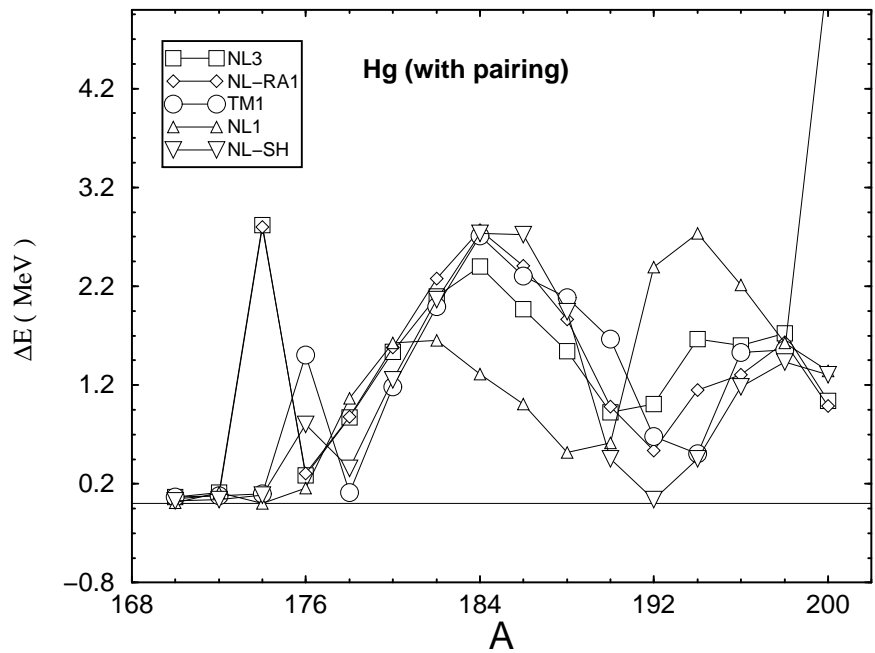
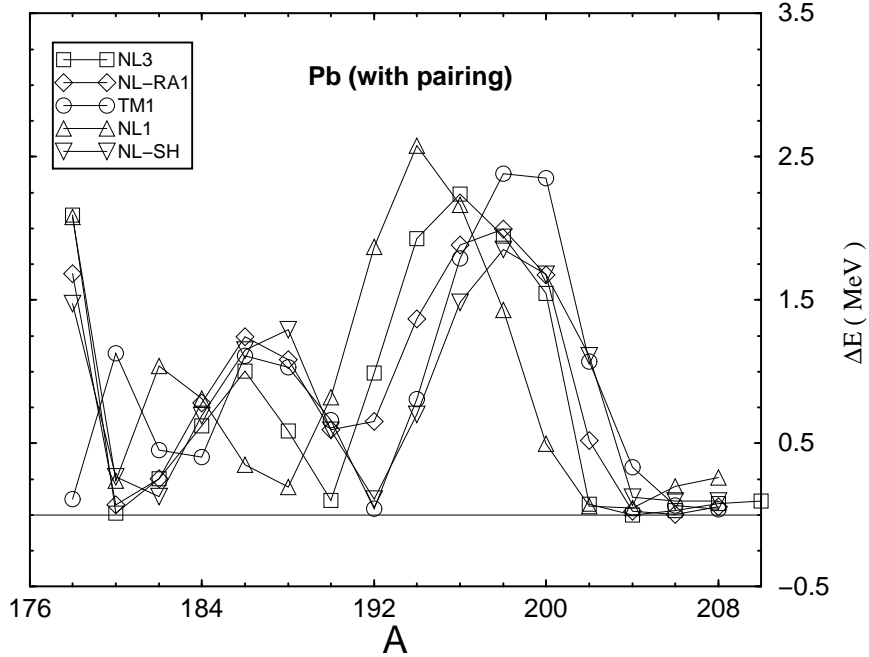
\noindent {\bf Figure 1}: The deformed RMF calculated difference 
           between the ground- and first excited-state binding 
	   energies as a function of the mass number A for various Hg and Pb 
	   isotopes, without (upper panel) and with (lower panel) pairing interaction.\\

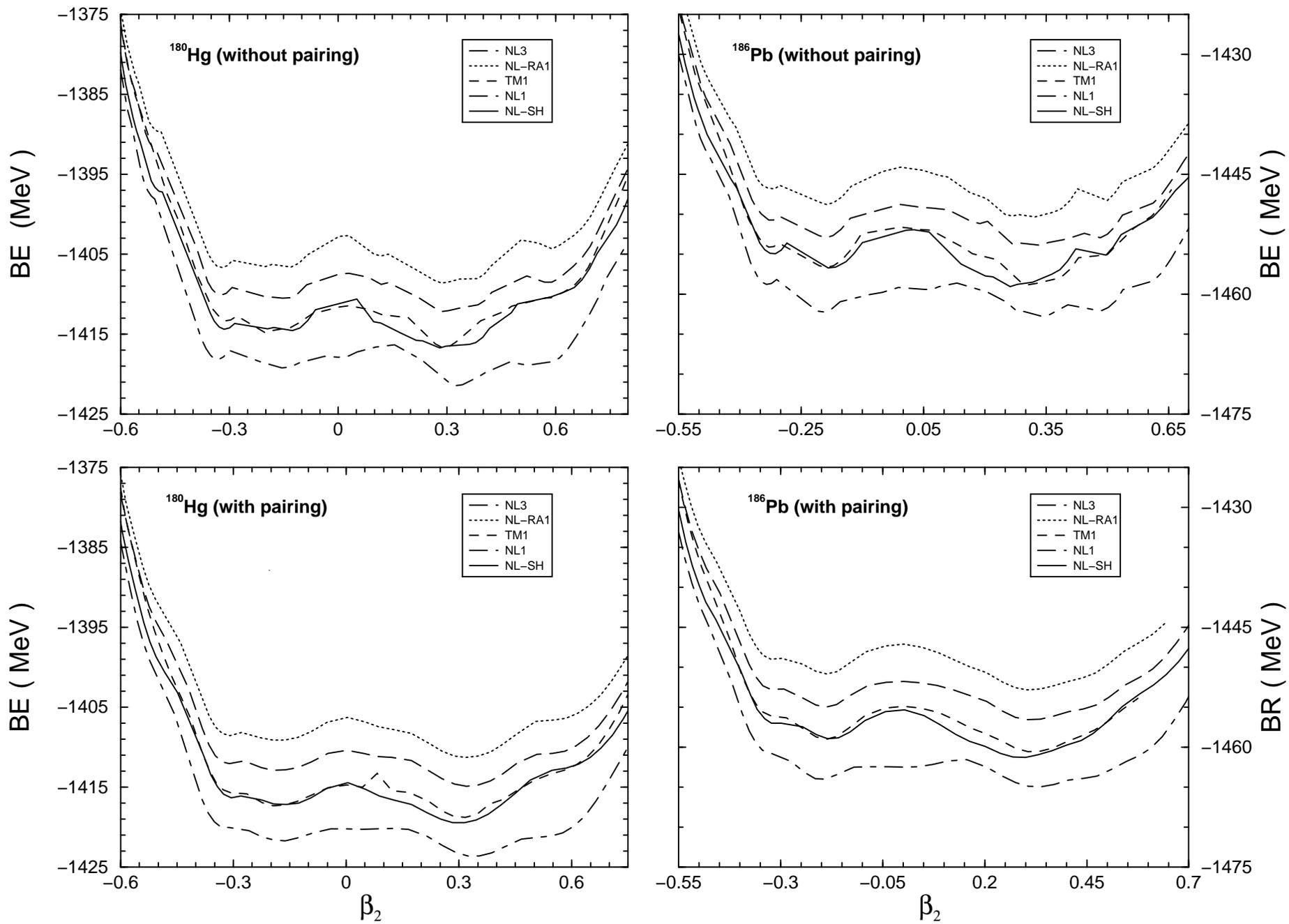
\noindent {\bf Figure 2}: 
          The deformed RMF calculated potential energy surfaces for the most 
          deformed $^{180}$Hg and $^{186}$Pb nuclei, both with pairing (upper panel) 
	  and without pairing (lower panel), using five parameter sets.\\

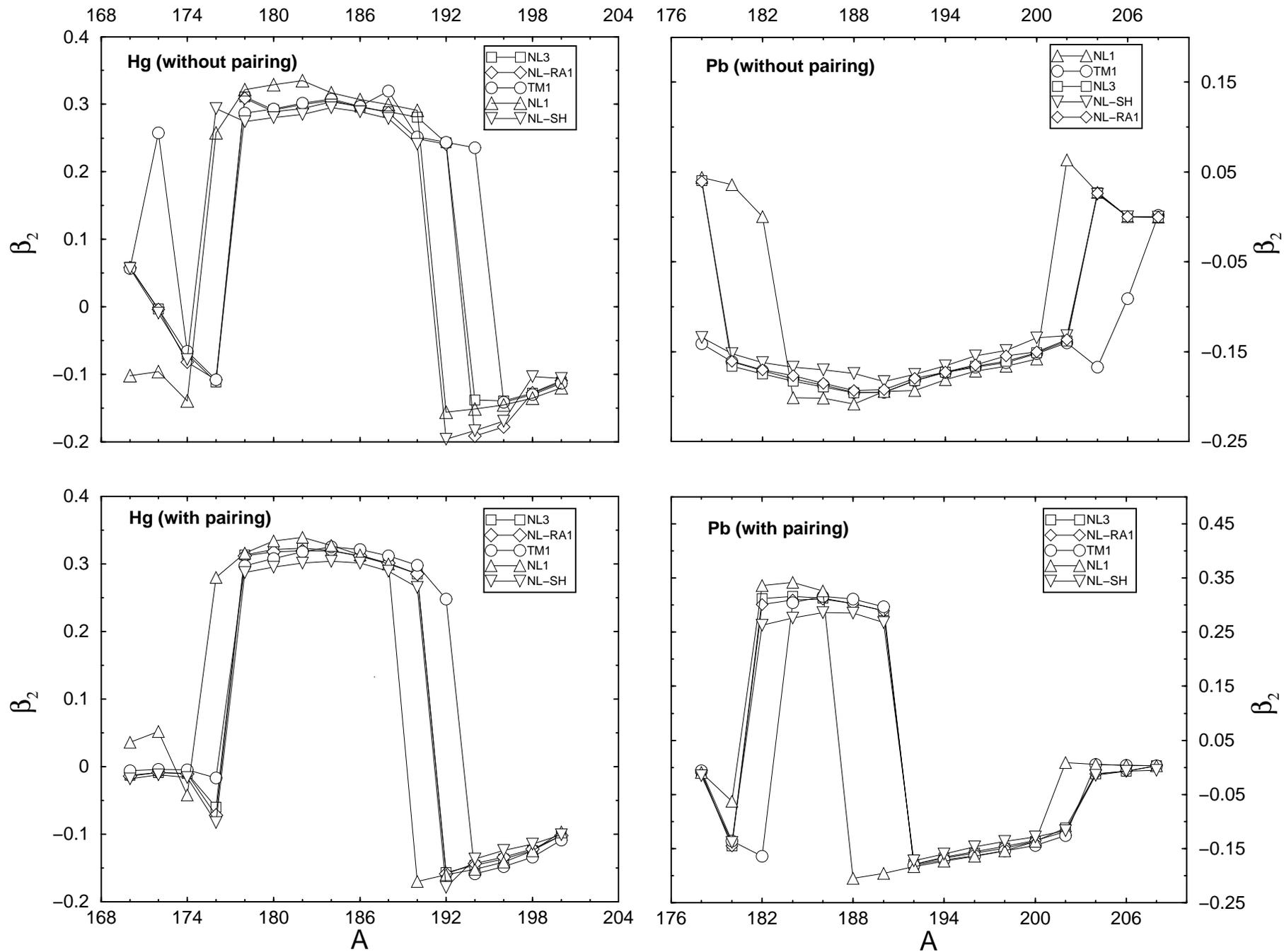
\noindent {\bf Figure 3}: The quadrupole deformation parameter as a function 
          of the mass number A for Hg and Pb nuclei, calculated on deformed RMF 
	  model with pairing (upper panel) and without pairing (lower panel) interactions.\\

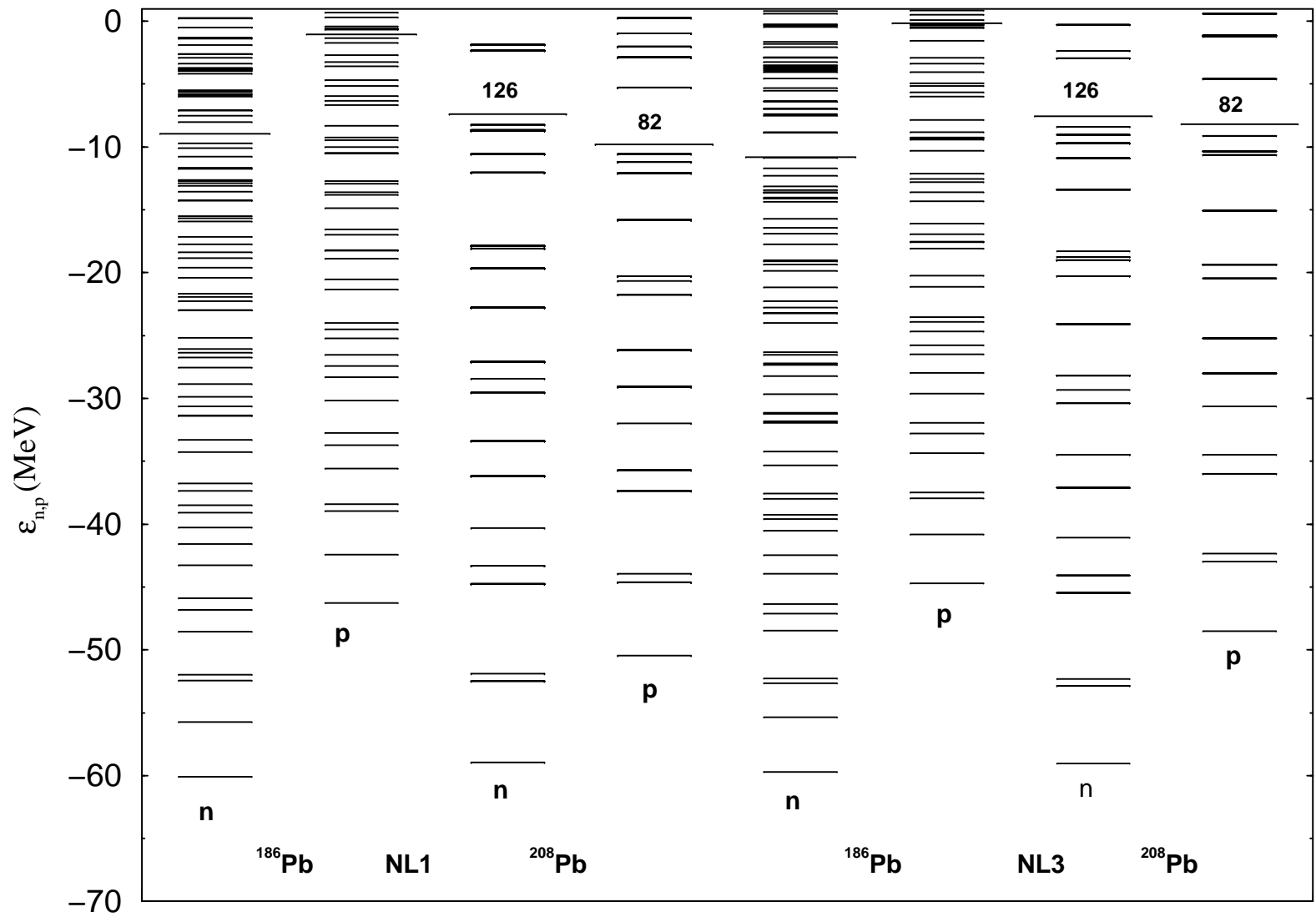
\noindent {\bf Figure 4}: The single-particle energy spectra for the 
          most spherical $^{208}$Pb and deformed $^{186}$Pb nuclei, calculated 
	  on deformed RMF model for NL1 and NL3 parameter sets. Pairing interaction is 
        included here.\\ 

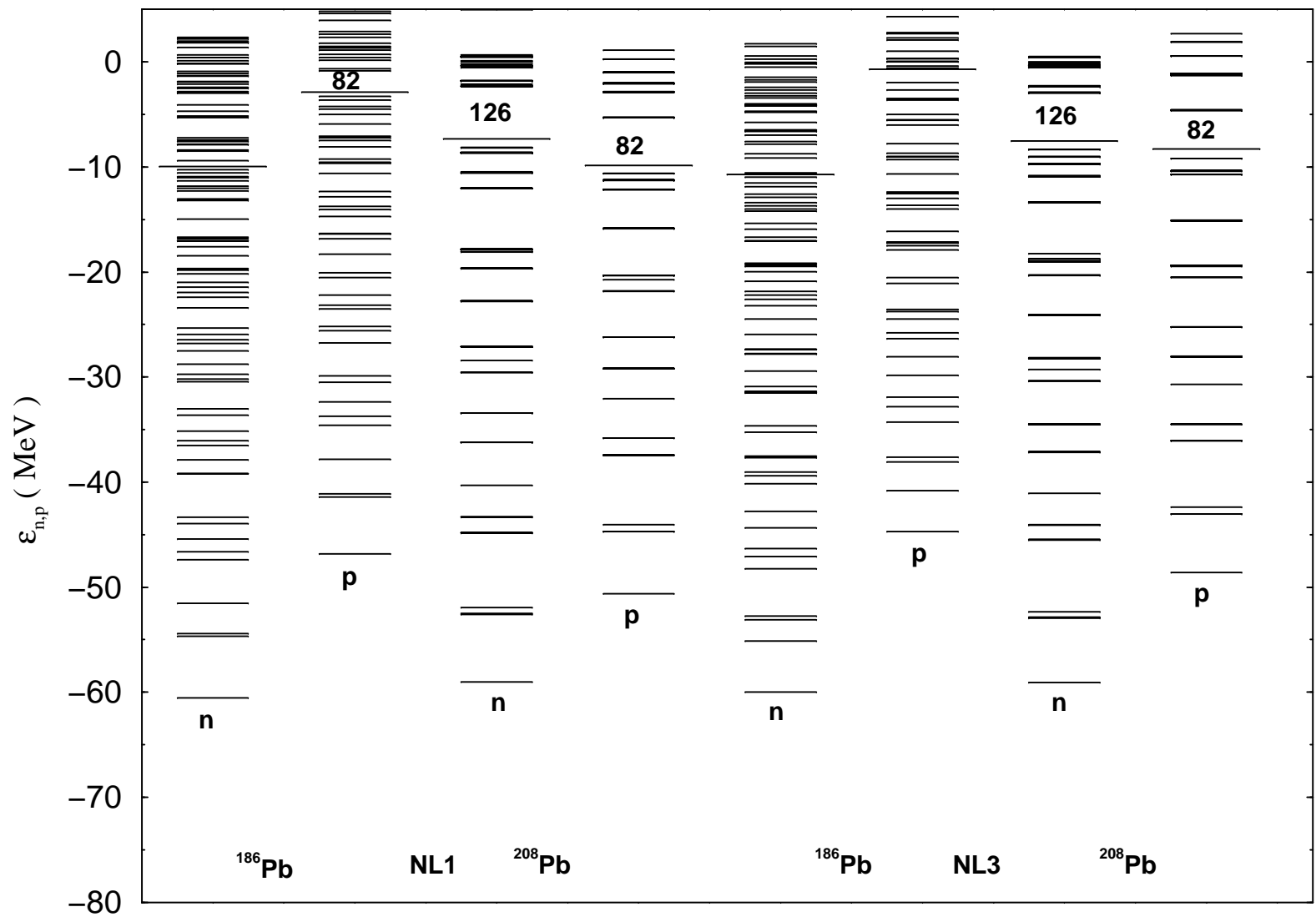
\noindent {\bf Figure 5}: Same as for Fig. 4 but for without pairing interaction.\\ 

\hfil\break
\hfil\break

\vfil
\eject

\vfil
\eject
\newpage
\oddsidemargin=-0.3in
\begin{table}
{Table 1: The ground-state binding energy BE and the quadrupole deformation 
parameter $\beta_2$ of some Hg and Pb isotopes for the five parameter sets, 
compared with experimental (or extrapolated) data \cite{audi97}. 
The binding energy is in MeV.}\\
\begin{small}
\begin{tabular}{|c|c|c|c|c|c|c|c|}
\hline
\hline 
\multicolumn{1}{|c}{Nuclei} 
&\multicolumn{1}{|c}{Sets}
&\multicolumn{1}{|c}{BE} 
&\multicolumn{1}{|c}{BE}
&\multicolumn{1}{c}{BE(expt.)} 
&\multicolumn{1}{c}{$\beta_{2}$}
&\multicolumn{1}{c|}{$\beta_{2}$}
&\multicolumn{1}{c|}{$\beta_{2}$(expt.)}\\
\hline
\hline
&&without pairing&with pairing&&without pairing&with pairing&\\
\hline
$^{184}$Hg&NL1&1458.7&1460.4&1448.7&0.317&0.327&0.156\cite{raman87}*\\
&NL3&1452.3&1453.9&&0.306&0.320&\\
&NL-SH&1456.8&1458.8&&0.295&0.304&\\
&NL-RA1&1448.8&1450.5&&0.304&0.319&\\
&TM1&1455.8&1457.6&&0.308&0.325&\\
$^{186}$Hg&NL1&1476.0&1478.4&1467.1&0.307&0.289&0.131\cite{raman87}*\\
&NL3&1469.6&1471.7&&0.296&0.311&\\
&NL-SH&1474.7&1476.9&&0.289&0.301&\\
&NL-RA1&1466.1&1468.4&&0.297&0.312&\\
&TM1&1472.6&1475.0&&0.297&0.321&\\
$^{196}$Hg&NL1&1554.1&1555.1&1551.3&-0.151&-0.139&0.116\cite{raman87}\\
&NL3&1550.9&1552.1&&-0.139&-0.137&\\
&NL-SH&1555.0&1556.6&&-0.117&-0.124&\\
&NL-RA1&1546.9&1548.3&&-0.137&-0.134&\\
&TM1&1550.1&1551.6&&-0.141&-0.148&\\
$^{198}$Hg&NL1&1567.8&1569.0&1566.5&-0.135&-0.120&0.107\cite{raman87}\\
&NL3&1566.3&1567.5&&-0.128&-0.124&\\
&NL-SH&1571.1&1572.3&&-0.103&-0.115&\\
&NL-RA1&1562.6&1564.0&&-0.127&-0.122&\\
&TM1&1564.8&1566.2&&-0.130&-0.134&\\
$^{200}$Hg&NL1&1581.4&1583.7&1581.2&-0.120&-0.096&0.098\cite{raman87}\\
&NL3&1581.8&1583.7&&-0.111&-0.109&\\
&NL-SH&1586.8&1588.6&&-0.106&-0.110&\\
&NL-RA1&1578.4&1580.3&&-0.109&-0.112&\\
&TM1&1579.45&1580.3&&-0.113&-0.109&\\
$^{190}$Pb&NL1&1498.9&1500.8&1489.7&-0.194&-0.188&-0.180\cite{dendoo89}\\
&NL3&1491.4&1493.2&&-0.195&0.289&\\
&NL-SH&1495.9&1497.0&&-0.183&0.268&\\
&NL-RA1&1487.4&1488.5&&-0.192&0.289&\\
&TM1&1493.9&1496.6&&-0.195&0.297&\\
$^{192}$Pb&NL1&1516.2&1518.0&1508.1&-0.193&-0.182&-0.175\cite{dendoo89}\\
&NL3&1509.2&1510.9&&-0.182&-0.173&\\
&NL-SH&1513.9&1515.0&&-0.175&-0.162&\\
&NL-RA1&1504.9&1506.4&&-0.180&-0.170&\\
&TM1&1511.3&1513.2&&-0.182&-0.181&\\
$^{194}$Pb&NL1&1533.2&1534.7&1526.0&-0.181&-0.175&-0.170\cite{dendoo89}\\
&NL3&1526.6&1528.3&&-0.173&-0.166&\\
&NL-SH&1531.0&1532.5&&-0.165&-0.156&\\
&NL-RA1&1522.4&1523.9&&-0.173&-0.164&\\
&TM1&1528.3&1530.1&&-0.173&-0.171&\\
\hline
\hline
\end{tabular}
* In \cite{bengt87a}, ${\beta_2}$=0.26, 0.25 respectively.
\end{small}
\end{table}
\newpage

\begin{table}
{Table 1: Continued...}\\ 
\begin{small}
\begin{tabular}{|c|c|c|c|c|c|c|c|c|c|}
\hline
\hline
\multicolumn{1}{|c}{Nuclei} 
&\multicolumn{1}{|c}{Sets}
&\multicolumn{1}{c}{BE} 
&\multicolumn{1}{c}{BE}
&\multicolumn{1}{c|}{BE(expt.)} 
&\multicolumn{1}{c|}{$\beta_{2}$}
&\multicolumn{1}{c|}{$\beta_{2}$}
&\multicolumn{1}{c|}{$\beta_{2}$(expt.)}\\
\hline
&&without pairing&with pairing&&without pairing&with pairing&\\
\hline
$^{204}$Pb&NL1&1610.8&1612.3&1607.5&0.027&0.000&0.041\cite{raman87}\\
&NL3&1609.1&1610.0&&0.027&0.000&\\
&NL-SH&1612.9&1613.9&&0.025&0.000&\\
&NL-RA1&1604.9&1606.0&&0.026&0.000&\\
&TM1&1604.8&1608.1&&0.027&0.006&\\
$^{206}$Pb&NL1&1627.0&1627.5&1622.4&0.000&0.000&0.032\cite{raman87}\\
&NL3&1625.4&1625.8&&0.000&0.000&\\
&NL-SH&1628.6&1629.1&&0.000&0.000&\\
&NL-RA1&1621.1&1621.6&&0.000&0.000&\\
&TM1&1619.5&1622.9&&0.051&0.004&\\
$^{208}$Pb&NL1&1641.6&1641.2&1636.5&0.000&0.000&0.054\cite{raman87}\\
&NL3&1640.5&1640.1&&0.000&0.000&\\
&NL-SH&1642.6&1642.7&&0.000&0.000&\\
&NL-RA1&1635.8&1635.8&&0.000&0.000&\\
&TM1&1636.3&1636.6&&0.002&0.003&\\
\hline
\hline
\end{tabular}
\end{small}
\end{table}
\begin{table}
\begin{small}
{Table 2: For $^{180}$Hg, the binding energy BE and quadrupole deformation 
parameter $\beta_2$ for normal and superdeformed bands, computed on deformed 
RMF model with and without pairing interaction, using various parameter sets.}
\end{small}
\begin{tabular}{|c|c|c|c|c|c|cc|}
\hline \hline
\multicolumn{1}{|c}{Nucleus}
&\multicolumn{1}{c}{Sets}
&\multicolumn{1}{c}{BE}
&\multicolumn{1}{c}{$\beta_2$}
&\multicolumn{1}{c}{BE}
&\multicolumn{1}{c}{$\beta_{2}$}
&\multicolumn{2}{c|}{shape}\\
\hline 
\multicolumn{2}{|c|}{} 
&\multicolumn{2}{|c}{with pairing}
&\multicolumn{2}{|c}{without pairing} 
&\multicolumn{2}{|c|}{}\\
\hline 
$^{180}$Hg&NL1&1422.405&0.323&1420.304&0.328&normal&\\
&&1420.420&0.569&1420.420&0.569&superdeformed&\\
&NL-SH&1420.811&0.307&1418.323&0.284&normal&\\
&&1411.157&0.955&1412.829&0.598&superdeformed&\\
&NL3&1414.833&0.317&1412.387&0.293&normal&\\
&&1410.907&0.572&1409.437&0.605&superdeformed&\\
&NL-RA1&1411.593&0.318&1409.189&0.291&normal&\\
&&1407.188&0.583&1405.702&0.606&superdeformed&\\
\hline\hline
\end{tabular}
\end{table}

\begin{thebibliography}{99}
\bibitem{faessler72} A. Faessler, U. G\"otz, B. Slavov, and T. Ledergerber, 
                     Phys. Lett. {\bf B39}, 579 (1972). 
\bibitem{calliau73} M. Cailliau, J. Leterssier, H. Flocard, and P. Quentin,
                     Phys. Lett. {\bf B46}, 11 (1973).  
\bibitem{kolb75} D. Kolb and C. Y. Wong, Nucl. Phys. {\bf A245}, 205 (1975).
\bibitem{bengtsson87} R. Bengtsson, et al., Phys. Lett. {\bf B183}, 1 (1987).
\bibitem{nazare93} W. Nazarewicz, Phys. Lett. {\bf B305}, 195 (1993).
\bibitem{patra94} S.K. Patra, S. Yoshida, and N. Takigawa, 
                  Phys. Rev. {\bf C50}, 1924 (1994).                
\bibitem{yoshida94} S. Yoshida, S.K. Patra, N. Takigawa, and C.R. Praharaj, 
                    Phys. Rev. {\bf C50}, 1398 (1994).
\bibitem{moeller88} D.G. Madland and J.R. Nix, Nucl. Phys. {\bf A476}, 1 (1988);
\par
P. M\"oller and J.R. Nix, At. Data Nucl. Data Tables {\bf 39}, 213 (1988).
\bibitem{heyde96} K. Heyde, C. De Coster, P. van Druppen, M. Huyse, J. L. 
                  Wood, and W. Nazarewicz, Phys. Rev. {\bf C53}, 1035 (1996).
\bibitem{taki96} N. Takigawa, S. Yoshida, K. Hagino, and S. K. Patra, 
                 Phys. Rev. {\bf C53}, 1038 (1996).
\bibitem{lala99} G.A. Lalazissis, S. Raman and P. Ring, At. Data and Nucl.
                 Data Tables, {\bf 71}, 1 (1999).
\bibitem{yoshida97} S. Yoshida and N. Takigawa, Phys. Rev. {\bf C55}, 1255 (1997).                 
\bibitem{niksic02} T. Nik\v si\'c, D. Vretenar, P. Ring, and G.A. Lalazissis,
                  Phys. Rev. {\bf C65}, 054320 (2002).
\bibitem{berger84} J.F. Berger, M. Girod, and D. Gogny, Nucl. Phys. {\bf A428}, 32 (1984).
\bibitem{novikov02} Yu. N. Novikov et al. Nucl. Phys. {\bf A697}, 92 (2002).  
\bibitem{toth99} K.S. Toth et al. Phys. Rev.  {\bf C60}, 011302(R) (1999). 
\bibitem{bender02} M. Bender, T. Cornelius, G.A. Lalazissis, J.A. Maruhn, W. Nazarewicz, and 
                   P.-G. Reinhard, Eur. Phys. J. {\bf A14}, 23 (2002). 
\bibitem{andreyev00} A.N. Andreyev, et al., Nature {\bf 405}, 430 (2000).
\bibitem{patra97} R.K. Gupta, S.K. Patra and W. Greiner, 
                  Mod. Phys. Lett. {\bf A12}, 1317 (1997);
\par 
                  S.K. Patra, R.K. Gupta and W. Greiner, 
                  Int. J. Mod. Phys. {\bf E6}, 641 (1997).
\bibitem{warner96} T.R. Werner, et al., Phys. Lett. {\bf B335}, 259 (1994);
\par 
		       T.R. Warner, et al., Nucl. Phys. {\bf A597}, 327 (1996).
\bibitem{patra00} S.K. Patra, W. Greiner and R.K. Gupta, 
                  J. Phys. G: Nucl. Part. Phys. {\bf 26}, L65 (2000); 
                  {\it ibid} {\bf 26}, 1569 (2000).
\bibitem{machleidt89} R. Machleidt, Adv. Nucl. Phys. {\bf 19}, 189 (1989).
\bibitem{patra91} S.K. Patra and C.R. Praharaj, Phys. Rev. {\bf C44},
                  2552 (1991).
\bibitem{patra01} M. Del  Estal, M. Centelles, X. Vi\~nas and S.K. Patra,
                  Phys. Rev. {\bf C63}, 024314 (2001).
\bibitem{rufa88} M. Rufa, P.-G. Reinhard, W. Greiner and M.R. Stranger,
                Phys. Rev. {\bf C38}, 390 (1988); 
\par
P.-G. Reinhard, Z. Phys. {\bf A329}, 257 (1993).
\bibitem{gambhir90} Y.K. Gambhir, P. Ring and A. Thimet, Ann. Phys. {\bf 198}, 132 (1990).
\bibitem{horowitz81} C.J. Horowitz and B.D. Serot, Nucl. Phys. {\bf A368}, 503 (1981).                   
\bibitem{serot86} B.D. Serot and J.D. Walecka, Adv. Nucl. Phys. {\bf 16}, 1 (1986).
\bibitem{rein86} P.-G. Reinhard, M. Rufa, J. Maruhn, W. Greiner, and J. Friedrich, 
                 Z. Phys. {\bf A 323}, 13 (1986).
\bibitem{sharma93} M.M. Sharma, M.A. Nagarajan, and P. Ring, 
                   Phys. Lett. {\bf B312}, 377 (1993). 
\bibitem{lala97} G.A. Lalazissis, J. K\"onig, and P. Ring, 
                 Phys. Rev. {\bf C55}, 540 (1997). 
\bibitem{sharma00} M.M. Sharma, A.R. Farhan, and S. Mythili, 
                   Phys. Rev. {\bf C61}, 054306 (2000). 
\bibitem{toki} Y. Sugahara and H. Toki, Nucl. Phys. A {\bf 579}, 557 (1994).
\bibitem{rash01} M. Rashdan, Phys. Rev. {\bf C63}, 044303 (2001).
\bibitem{raman87} S. Raman, et al., At. Data Nucl. Data Tables {\bf 36}, 1 (1987).
\bibitem{audi97} G. Audi, O. Berstllon, J. Blachot and A.H. Webpstra, 
                 Nucl. Phys. {\bf A624}, 1 (1997).
\bibitem{dendoo89} P. Dendooven, et al., Phys. Lett. {\bf B226}, 27 (1989).                  
\bibitem{bengt87a} R. Bengtsson and W. Nazarewicz, Lund-Mph-87/08 preprint.  
\bibitem{frauendorf75} S. Frauendorf and V.V. Pashkevich, Phys. Lett. {\bf B55}, 365 (1975).  
\bibitem{dracolis88} G.D. Dracoulis, et al., Phys. Rev. {\bf B208}, 365 (1988).                  
\bibitem{patra02} S.K. Patra, B.K. Raj, M.S. Mehta and R.K. Gupta, 
                 Phys. Rev. {\bf C65}, 054323 (2002).
\bibitem{julin01} R. Julin, et al., J. Phys. G: Nucl. Part. Phys. {\bf 27}, R109 (2001).

\end{thebibliography}
\end{document}